\documentclass[epj,twocolumn]{webofc}
\usepackage[varg]{txfonts}   % Web of Conferences font
\woctitle{New Frontiers in Physics 2014}

\usepackage{tikz}
\usetikzlibrary{snakes}
\begin{document}
\title{Search for a dileptonic edge with CMS}

\author{Konstantinos Theofilatos (on behalf of the CMS Collaboration)\inst{1}\fnsep\thanks{\email{konstantinos.theofilatos@cern.ch}} 
}

\institute{Institute for Particle Physics, ETH Z\"urich, R\"amistrasse 101, 8092 Z\"urich,Switzerland}

\abstract{
We present a search for a kinematic edge in the invariant mass distribution of two opposite-sign same-flavor leptons, 
in final states with jets and missing transverse energy. 
The analysis makes use of $19.4$~fb$^{-1}$ proton-proton collision data at $\sqrt{s} = 8$~TeV. The data have been recorded with the CMS experiment.
Complementary methods have been used for the background estimation, which when combined achieve a total uncertainty of $5\%$ ($10\%$) for leptons in the central (forward) rapidity
of the detector.
We do not observe a statistically significant signal  and the results are consistent with the background-only hypothesis.
}
\maketitle

A search for new physics in final states with an opposite-sign same-flavor lepton pair (dilepton), jets and missing transverse energy (MET) is presented here.
We search for evidence of production of a new heavy neutral particle, which is produced in the cascade of a strongly interacting heavier resonance.
This report has been compiled from results that have been published here~\cite{PAS}.
We use proton-proton collision data at $\sqrt{s} = 8$~TeV. The total accumulated luminosity is $19.4$~fb$^{-1}$. 
The data have been recorded with the CMS experiment~\cite{CMS}.

\begin{figure}[htbp]
\begin{flushleft}                           
 %\begin{tikzpicture}[snake=zigzag, line before snake = 5mm, line after snake = 5mm, scale = 1.15]
 \begin{tikzpicture}[scale = 1.15]
 %draw horizontal line   
 \draw[->] (0.0,0) -- (6.0,0);

 \draw[->] (0.0,0) -- (1.5,1.5);
 \draw (1.5,1.5) node[below=3pt] {$  $} node[above=3pt] {\Large $q$};

 \draw[->] (2,0) -- (3.5,1.5);
 \draw (3.5,1.5) node[below=3pt] {$  $} node[above=3pt] {\Large $ \ell^{\pm} $};

 \draw[->] (4,0) -- (5.5,1.5);
 \draw (5.5,1.5) node[below=3pt] {$  $} node[above=3pt] {\Large $ \ell^{\mp} $};

 \filldraw (0,0) circle [radius=0.1, fill=cyan, draw=blue];
 \draw (0,0) node[below=3pt] {\Large $ \tilde{q} $} node[above=3pt] {$  $};
 \filldraw (2,0) circle [radius=0.1, fill=cyan, draw=blue];
 \draw (2,0) node[below=3pt] {\Large$ \tilde{\chi}^{0}_{2} $} node[above=3pt] {$   $};
 \filldraw (4,0) circle [radius=0.1, fill=cyan, draw=blue];
 \draw (4,0) node[below=3pt] {\Large$ \tilde{\ell}^{\mp} $} node[above=3pt] {$   $};
 \draw (6.0,0) node[below=3pt] {\Large $\tilde{\chi}^{0}_{1}  $} node[above=3pt] {invisible};
% \filldraw (6,0) circle [radius=0.1];
 \end{tikzpicture}
\end{flushleft}                           
 \caption{An example SUSY process that will lead to an excess of events in the invariant mass distribution of the two leptons $m_{{\ell}{\ell}}$. The invariant mass
distribution of the two leptons will exhibit a threshold effect reaching a maximum endpoint.
The location of the endpoint solely depends on the masses of the particles involved and can be used to constrain different physics models.}
 \label{fig:cart}
\end{figure}
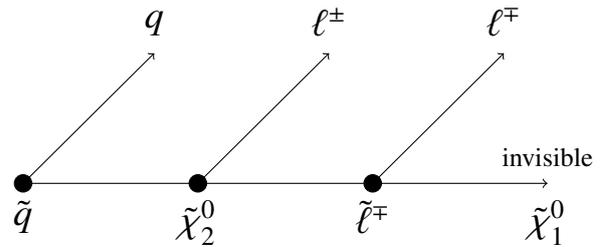

The invariant mass distribution of the dilepton system can exhibit a rise that increases with dilepton mass followed by a sharp cut-off, 
resembling the shape of a triangle with an edge, if the two leptons originate from the decay of a heavy neutral particle.
In Supersymmetry (SUSY), such a signal is exemplified by the decay of the heavy neutralino $(\tilde{\chi}^{2}_{0})$, 
which is produced within the cascade of a squark ($\tilde{q}$) -- the SUSY partner of a quark. 
An example cascade is depicted in Fig.~\ref{fig:cart}.
The cascade can also start from a gluino, the SUSY partner of the gluon, if that is heavier than the squark resulting to an
additional energetic jet in the final state. 
In either cases, the location of the dileptonic mass endpoint is given by Eq.~\ref{eq1}.
 
{\large
\begin{equation}
\label{eq1}
m_{{\ell}{\ell}}^{max} = m_{\tilde{\chi}^{0}_{2}} \sqrt{
\left(    
1 - \left(\frac{m_{\tilde{\ell}}}{m_{\tilde{\chi}^{0}_{2}}}\right)^2 
\right)
\left(    
1 - \left(\frac{m_{\tilde{\chi}^{0}_{1}}}{m_{\tilde{\ell}}}\right)^2 
\right)
}
\end{equation}
}

If the mass difference between the two neutralinos is small, or the mass of the slepton $(\tilde{\ell})$ is large, 
the  $\tilde{\chi}^{2}_{0}$ decay can proceed through a virtual $\mathrm{Z}^{0*}$ boson exchange, resulting to a direct 3-body decay.
The endpoint in that case is given by the mass difference of the two neutralinos, $m_{{\ell}{\ell}}^{max} = m_{\tilde{\chi}^{0}_{2}} - m_{\tilde{\chi}^{0}_{1}}$
and the invariant mass distribution of the dilepton system is not anymore an exact triangle and it has more rounded shape.
If there is sufficient energy to produce a $\mathrm{Z}^0$ boson, i.e. when $m_{\tilde{\chi}^{0}_{2}} - m_{\tilde{\chi}^{0}_{1}} \sim m_\mathrm{Z}$ 
this will be preferred against a possible 3-body decay and the dilepton mass distribution will exhibit an excess of events at $m_\mathrm{Z}\sim91$~GeV. 
Furthermore, similar cascades involving higher mass neutralino eigenstates $m_{\tilde{\chi}^{0}_{x = 2,3,4}}$ decaying to lighter neutralinos can occur 
producing similar topologies. 

Measuring the possible different endpoints and their $\sigma \times \mathrm{BR}$ is a way to constrain candidate new physics models with data~\cite{edge1}.
In addition, since the above discussion was only based on kinematic argumentation, any massive heavy neutral particle (SUSY or not) that decays through the 
aforementioned cascades may give rise to similar phenomenology.

In SUSY, the lightest neutralino $\tilde{\chi}^{0}_{1}$ is massive, neutral, weakly interacting and is most often assumed to be stable. 
This assumption is not a necessity, but is experimentally well motivated and can explain the evidence of dark matter in the universe.
The stability of dark matter particles in the universe implies the conservation of 
a new quantum number and thus the dark matter particles are expected to be always pair produced in a collider,
if there is sufficient $\hat{s}$ available.
The stability of $\tilde{\chi}^{0}_{1}$ can be enforced in SUSY by the R-parity conservation~\cite{susyprimer}. 
Being a carbon copy of the SM, we expect that SUSY particles will be predominantly produced via the strong interaction $pp \to \tilde{q}\tilde{q}$, $\tilde{g}\tilde{q}$ and $\tilde{g}\tilde{g}$. In each signal event, two independent cascades will be formed, originating from the decays of the squarks and gluinos
that will decay producing jets and gauge bosons until the lightest SUSY particle is produced $\tilde{\chi}^{0}_{1}$.
We therefore expect at least two $\tilde{\chi}^{0}_{1}$ in the final state. 
A stable $\tilde{\chi}^{0}_{1}$ will escape from detection like a heavy neutrino, producing missing transverse energy in the detector.

Both CMS and ATLAS Collaborations have extensively studied the dilepton signature as a search strategy before the LHC start-up~\cite{ptdr2}\cite{atlasptdr}.
Previous results of a search for kinematic edge have been published by CMS at $\sqrt{s} = 7$~TeV~\cite{edge2011}.
In addition, a search for a $\mathrm{Z}^0$ boson accompanied by jets and MET has been also published by CMS at $\sqrt{s} = 7$~TeV~\cite{jzbpaper}.
This update, uses $\sqrt{s} = 8$~TeV data and unifies the experimental methods deployed earlier in the corresponding CMS $7$~TeV papers ~\cite{edge2011}\cite{jzbpaper}. 

\begin{figure}
\centering
\includegraphics[width=8cm,clip]{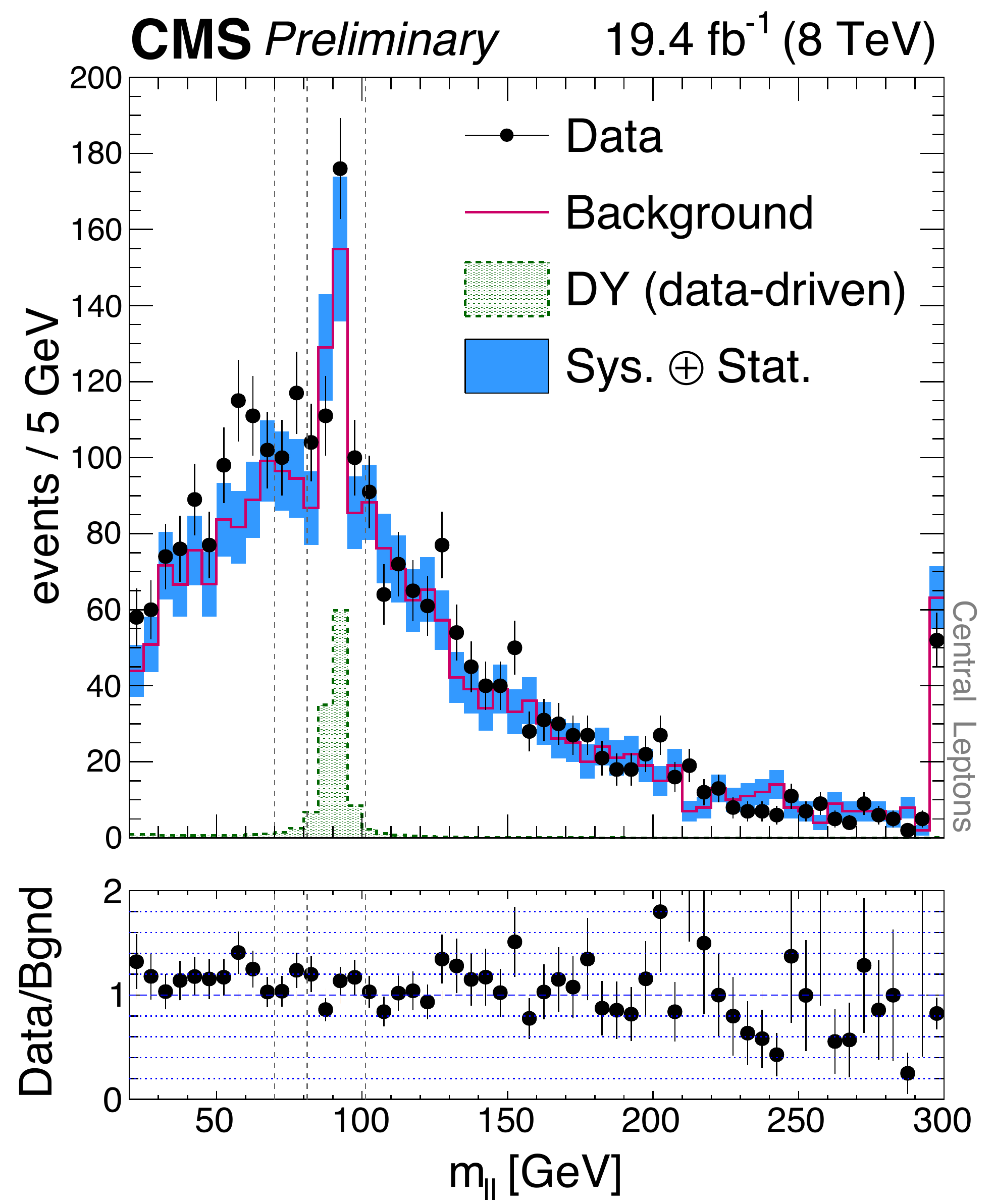}
\caption{
Comparison between the observed and estimated SM background dilepton mass distributions in the central region, 
where the SM backgrounds are evaluated from control samples in data.
}
\label{fig8}       
\end{figure}

\begin{figure}
\centering
\includegraphics[width=8cm,clip]{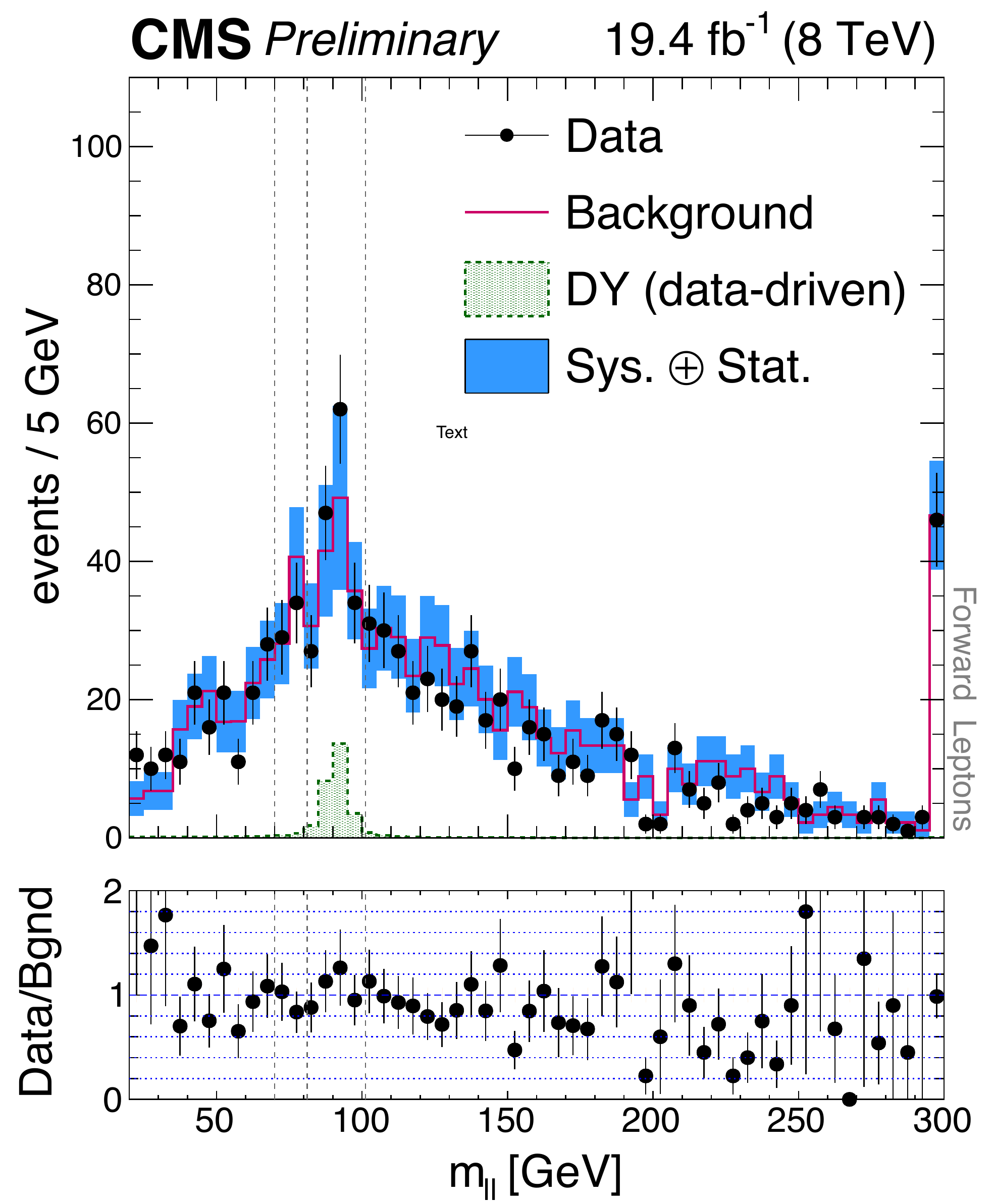}
\caption{
Comparison between the observed and estimated SM background dilepton mass distributions in the forward region, 
where the SM backgrounds are evaluated from control samples in data.
}
\label{fig9}       % Give a unique label
\end{figure}

We select events with two leptons having $p_\mathrm{T} > 20$~GeV and $|\eta|<2.4$, with $p_\mathrm{T}$ denoting the transverse momentum and $\eta$ the pseudorapidity.
We also count the number of jets produced in the event having $p_\mathrm{T} > 40$~GeV and $|\eta|<3.0$.
In the jet counting, we exclude jets that are within $\Delta R = \sqrt{\delta\phi^2 + \delta\eta^2}<0.4$ away from the leptons. 
We require that the two leptons are well separated ($\Delta R>0.3$).
The signal region is defined as ($N_\mathrm{jets} = 2$ $\cup$ $\mathrm{MET}> 150$~GeV) $\cap$ ($N_\mathrm{jets} \geq 3$ $\cup$ $\mathrm{MET}>100$~GeV),
and the phase space is further subdivided to ``central'' region (Fig.~\ref{fig8}) where both leptons have $|\eta|<1.4$ and ``forward'' region when 
at least one lepton lays in the forward region $1.6 < |\eta|< 2.4$ (Fig.~\ref{fig9}). 

A priori these two regions of the phase space, have different $S/B$. 
We expect that the leptons that will originate from the decay of heavy resonance, 
will be more centrally distributed in rapidity with respect to the standard model (SM) background processes that mimic the same topology. 
The main SM background is top pair production, 
\begin{equation}
t\bar{t} \to W^+W^- b\bar{b} \to \ell^{+} \ell^{-}  b\bar{b} \nu \bar{\nu}
\end{equation}
which results in equal yields of same-flavor (SF = $ee$ and $\mu\mu$) and opposite-flavor (OF = $e\mu$) events.
This is due to the lepton universality in the decays of the two $\mathrm{W}$ bosons, which decay with equal probability to all lepton flavors.
We use a control sample of OF data that satisfies all signal selection criteria, but the flavor composition which is SF for the signal and OF for the control sample.
The OF data are corrected for triggering and lepton reconstruction efficiency effects such as to match the SF selection efficiency. 
The measured multiplicative correction factor $\text{R(SF/OF)} = N_\mathrm{SF}/N_\mathrm{OF}$ is consistent with unit 
within uncertainties and is very robust against changes in the 
muon over electron efficiency ratio $(r=\epsilon_{\mu}/\epsilon_{e})$.
This can be understood by considering that $\text{R(SF/OF)} = 0.5 \times (r + r^{-1})\times R_\mathrm{T}$ with $R_\mathrm{T}$ a constant factor related to the triggering 
efficiencies of $ee$, $\mu\mu$ and $e\mu$ paths when they reach their plateau.
It is easy to verify that  $0.5 \times (r + r^{-1})$ stays close to $1.0$  even when $r$ varies significantly, due to cancellation efficiency effects when combining electron
and muons channels to form the SF sample. 
We measure the efficiency correction factors and their uncertainty, by two complementary methods, which are detailed in~\cite{PAS}.
The results of the two methods are consistent and the two methods are finally combined.
The final uncertainty of $\text{R(SF/OF)}$ is 4\% (7\%) for the central (forward) region.
Another important source of background comes from Drel--Yan (DY) events, which could exhibit large apparent missing energy due to detector resolution and reconstruction effects.
This background is not flavor symmetric and therefore the OF control sample is not alone sufficient to cope with it.
We deploy two independent methods to estimate the DY background, the MET templates and the Jet-Z-Balance (JZB) method.
The former uses photon+MET data reweighed to match the DY kinematics while the latter exploits the sign symmetry of the JZB variable,
which is an estimate of the momentum imbalance in DY events~\cite{jzbpaper}.
The total uncertainty on the background is 5\% (10\%) for the central (forward) region.
The region of phase space with moderate MET and hadronic activity HT\footnote{HT is defined as the scalar sum of $p_\mathrm{T}$ of all jets} that is probed by this analysis, 
is not easily accessible to inclusive searches for jets+MET that use MET and HT triggers.
The achieved level of precision allows to search for soft signals that would have been missed in more inclusive ``all-hadronic'' final states, which  
typically suffer from much higher SM background uncertainties.

\begin{figure}
\centering
\includegraphics[width=8cm,clip]{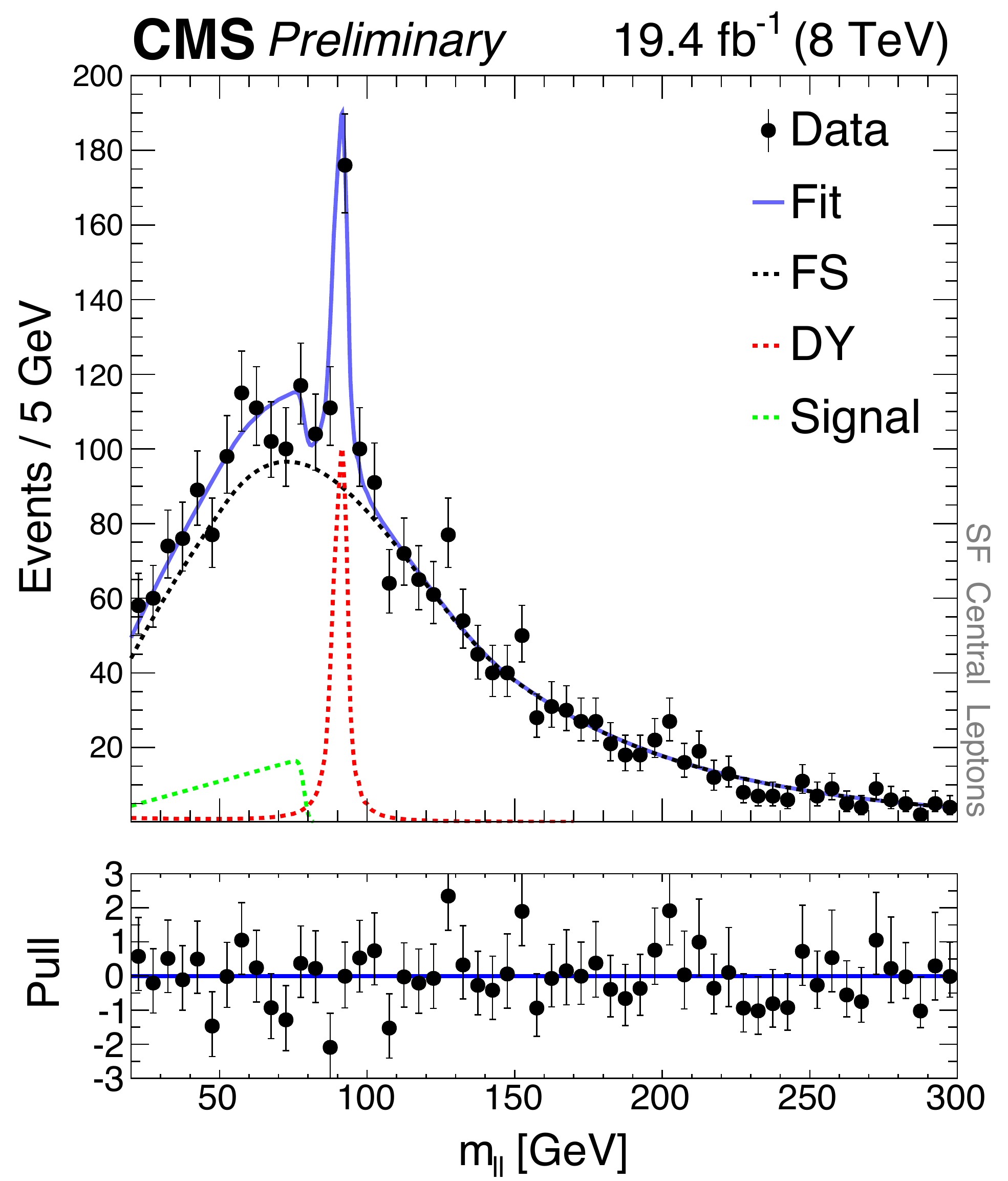}
\caption{ 
Fit results for the signal-plus-background hypothesis in comparison to the measured dilepton mass distributions, in the central region, projected on the SF event sample. 
The combined fit shape is shown as a blue, solid line. 
The individual fit components are indicated by dashed lines. 
The flavor-symmetric background is denoted as FS and is displayed with a black dashed line. 
The Drell--Yan contribution is denoted as DY and is displayed with a red dashed line. 
The extracted signal component is denoted as Signal and is displayed with a green dashed line.}
\label{fig3}       % Give a unique label
\end{figure}
\begin{table}[hbtp]
 \renewcommand{\arraystretch}{1.3}
 \setlength{\belowcaptionskip}{6pt}
% \small
\caption{Results of the counting experiment.}

 \centering
  \label{tab:METresults2012}

  \begin{tabular}{l| cc}
    \hline
    \hline
                           &  Central          & Forward        \\
    \hline
     Observed data events $(N_\mathrm{data})$ &  860               & 163             \\
    \hline
     Estimated background $(N_\mathrm{b})$   & $730\pm40$         & $157\pm16$      \\
    \hline
    $N_\mathrm{s} = N_\mathrm{data} - N_\mathrm{b}$  & $130^{+48}_{-49}$  & $6^{+20}_{-21}$ \\
    \hline
     Significance [$\sigma$]   &2.6  &  $0.3$    \\
   \hline
    \hline
  \end{tabular}
\end{table}

\begin{figure}
\centering
\includegraphics[width=8cm,clip]{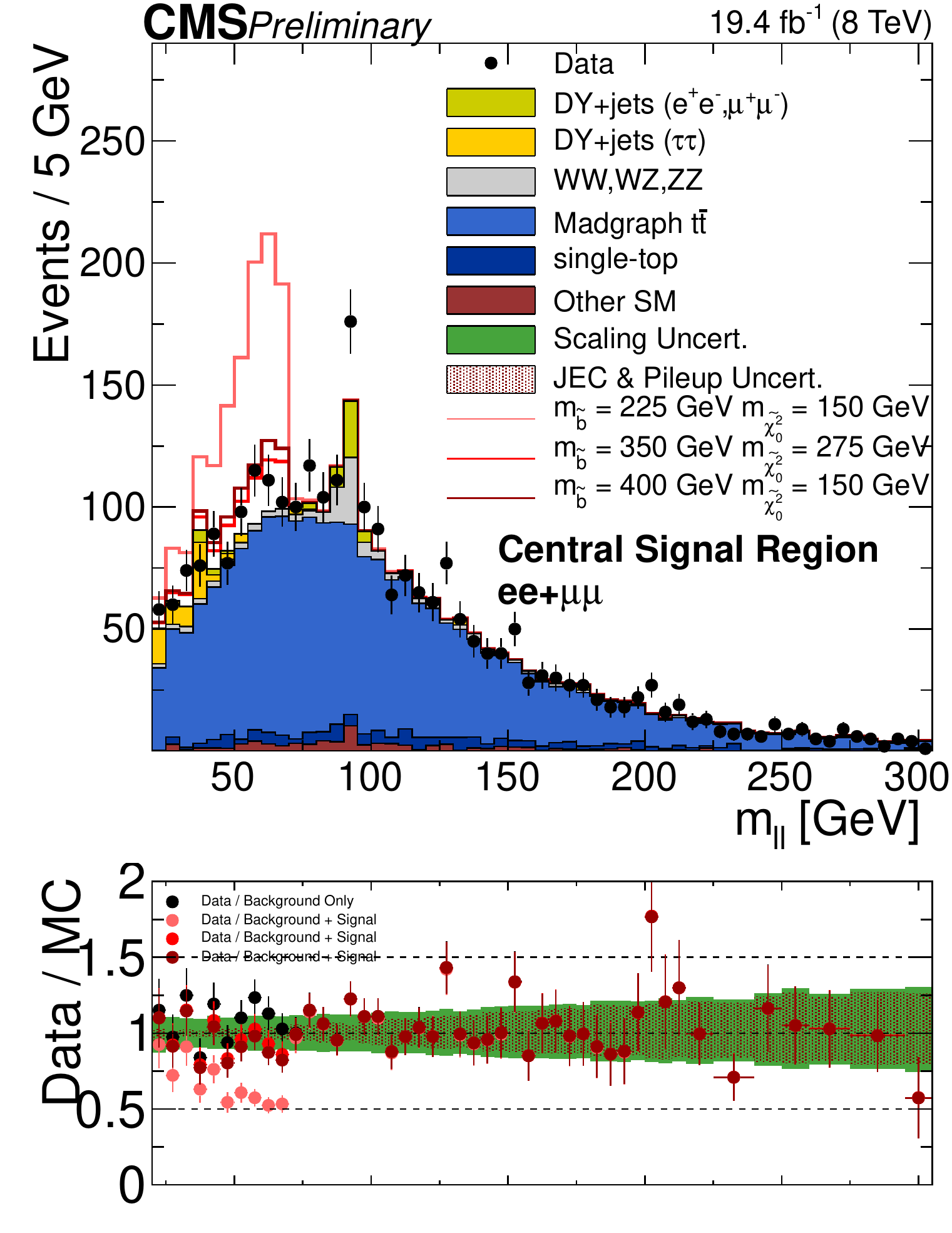}
\caption{Data compared with SM simulation in the central signal region. Example signal scenarios based on the pair production of bottom-squarks is also overlaid.}
\label{fig12}       % Give a unique label
\end{figure}

A counting experiment is performed for $20 <m_{\ell\ell} < 70$~GeV while the search for kinematic edge is performed in $20 <m_{\ell\ell} < 300$~GeV by fitting 
the signal and background shapes in data (Fig.~\ref{fig3}). 
The background shape normalization is performed with a Gaussian constraint taking into account the measured value of $\text{R(SF/OF)}$ and its uncertainty.
For the signal, we use a triangular shape assumption, which is the nominal shape expected from kinematics for 
two sequential two-body decays as e.g. in SUSY when the decay occurs via an intermediate slepton.
The local significance of the counting experiment 
is $2.6\sigma$ $(0.3\sigma)$ in the central (forward) signal region.
The best fitted value for the location of the kinematic edge, assuming a triangular signal shape, is at $78.7 \pm 1.4$~GeV (Fig.~\ref{fig3}), with significance of $2.4\sigma$.
The fit is simultaneous to both central and forward signal regions. 
The reduced significance of the fit can be understood by the mixing with equal weight
of the central and forward regions.
The comparison of data versus simulation of the background and three different benchmark models is shown in Fig.~\ref{fig12}. 
These reference signal models are based on sbottom pair production with masses as indicated in the legend. 
The decay mode of each sbottom quark is $\tilde{b} \to \tilde{\chi}_2^0 b \to \tilde{\chi}_1^0 \mathrm{Z}^{0*} b$
with $\mathrm{Z}^{0*}$ boson being off-shell and decaying to leptons, quark--antiquark and invisible with their SM branching fractions.
As a result, about 13\% of the signal events contain at least one SF dilepton in the final state. 
The mass difference $m_{\tilde{\chi}_2^0} - m_{\tilde{\chi}_1^0}$ has been fixed to $70$~GeV, producing the characteristic endpoint expected for a 3-body decay.

The results of the counting experiment are summarized in Tab.~\ref{tab:METresults2012}.
In the forward signal region, which is experimentally more challenging, 
we record as much data as expected from the SM background.
In the central signal region, although the yield of signal candidates $N_\mathrm{s} = 130^{+48}_{-49}$ is sizable, it is not 
sufficiently large to reject the background-only hypothesis. The possibility that the observed excess is solely a background fluctuation of statistical nature,
can not be excluded. 
This analysis probes a phase space that is not easily accessible to inclusive searches for jets+MET and it does that with foremost precision.
We will have to wait for its repetition with 13~TeV LHC data in order to wipe out the possibility that this is a first hint of new physics.

%\clearpage 

\end{document}